\documentclass[unnumsec,webpdf,contemporary,large]{oup-authoring-template}%

\graphicspath{{Fig/}}

\theoremstyle{thmstyleone}%
\theoremstyle{thmstyletwo}%
\theoremstyle{thmstylethree}%

\def\OURS{XATGRN}
\usepackage{amsmath}
\usepackage{booktabs}
\usepackage[most]{tcolorbox}

\begin{document}

\journaltitle{Journal Title Here}
\DOI{DOI HERE}
\copyrightyear{2022}
\pubyear{2019}
\access{Advance Access Publication Date: Day Month Year}
\appnotes{Paper}

\firstpage{1}


\title[Short Article Title]{Cross-Attention Graph Neural Networks for Inferring Gene Regulatory Networks with Skewed Degree Distribution}

\author[2]{Jiaqi Xiong$^\dagger$}
\author[3]{Nan Yin$^\ast$$^\dagger$}
\author[4]{Shiyang Liang$^\dagger$\ORCID{0000-0002-5357-5145}}
\author[2]{Haoyang Li}
\author[5]{Yingxu Wang}
\author[6]{Duo Ai}
\author[4]{Fang Pan}
\author[1]{Jingjie Wang$^\ast$}

\authormark{Author Name et al.}
\address[1]{\orgdiv{Department of Gastroenterology}, \orgname{Tangdu Hospital, Fourth Military
Medical University}, \orgaddress{\postcode{710038}, \state{Shaanxi}, \country{China}}}

\address[2]{\orgdiv{Aberdeen Institute of Data Science and Artificial Intelligence}, \orgname{South China Normal University}, \orgaddress{\postcode{528225}, \state{Guangzhou}, \country{China}}}
\address[3]{\orgdiv{Department of Computer Science and Engineering}, \orgname{Hong Kong University of Science and Technology }, \orgaddress{\state{Hong Kong}, \country{China}}}
\address[4]{\orgdiv{Department of Internal Medicine}, \orgname{The No. 944 Hospital of Joint Logistic Support Force of PLA}, \orgaddress{\street{Xiongguan Road}, \postcode{735000}, \state{Jiu Quan}, \country{China}}}
\address[5]{\orgdiv{Department of Machine Learning}, \orgname{Mohamed bin Zayed University of Artificial Intelligence}, \orgaddress{\state{Abu Dhabi}, \country{UAE}}}
\address[6]{\orgdiv{Department of Dermatology}, \orgname{Xijing Hospital, Fourth Military Medical University,} \orgaddress{\street{  No 127 of West Changle Road}, \postcode{710032}, \state{Xi'an, Shaanxi}, \country{China}}}

\corresp[$\ast$]{Corresponding author. \href{email:shiyang.liang@outlook.com}{yinnan8911@gmail.com
; jingjiefmmu@163.com}}

\received{Date}{0}{Year}
\revised{Date}{0}{Year}
\accepted{Date}{0}{Year}



\abstract{Inferencing Gene Regulatory Networks (GRNs) from gene expression data is a pivotal challenge in systems biology, and several innovative computational methods have been introduced. However, most of these studies have not considered the skewed degree distribution of genes. Specifically, some genes may regulate multiple target genes while some genes may be regulated by multiple regulator genes. Such a skewed degree distribution issue significantly complicates the application of directed graph embedding methods. To tackle this issue, we propose the Cross-Attention Complex Dual Graph Embedding Model (XATGRN). Our XATGRN employs a cross-attention mechanism to effectively capture intricate gene interactions from gene expression profiles. Additionally, it uses a Dual Complex Graph Embedding approach to manage the skewed degree distribution, thereby ensuring precise prediction of regulatory relationships and their directionality. Our model consistently outperforms existing state-of-the-art methods across various datasets, underscoring its efficacy in elucidating complex gene regulatory mechanisms. Our codes used in this paper are publicly available at: https://github.com/kikixiong/XATGRN.}


\maketitle

\section{Introduction}
Cells execute diverse functions by expressing various genes and through the interplay of regulatory relationships among these genes. Bulk sequencing~\cite{janjic2022prime} enables the profiling of gene expression within specific tissues. From such gene expression matrices, researchers can extract a Gene Regulator Network (GRN). 
GRNs are crucial in different biological processes; they can aid in studying developmental biology, unraveling the mechanisms behind various diseases, and identifying new therapeutic targets~\cite{levine2005gene,dong2024deep,kloesch2022gata6,meng2024heterogeneous}.

Although numerous databases have accumulated extensive regulatory relationships, each tissue's GRN reveals its unique regulatory characteristics~\cite{goldman2020gene,sonawane2017understanding,yi2024bpp}. Hence, it is impractical to validate the specific GRN regulatory network in each tissue solely through wet experiments. Many researchers have proposed computational methods for bulk sequencing GRN inference. Prior to the prevalent of deep learning models, researchers proposed various GRN inference methods based on conventional machine learning and statistical approaches, including correlation-based methods~\cite{salleh2015reconstructing}, Bayesian network-based methods~\cite{liu2016inference,xing2017improved}, and hybrid methods~\cite{raza2016recurrent,ju2024survey}. In recent years, with the accumulation of sequencing data~\cite{katz2022sequence,shiraishi2022systematic} and the development of deep learning technology, numerous deep learning-based bulk GRN inference methods have emerged. For instance, the CNNGRN~\cite{gao2023cnngrn} model utilizes a convolutional neural network to reconstruct gene regulatory networks from large-scale gene expression data. This CNNGRN model leverages known gene regulatory networks as prior knowledge to capture gene neighborhood information, and incorporates it as network structural features to enhance the predictive power over gene-gene relations. This improvement is particularly effective in inferring gene regulatory networks in real species.

Although effective, CNN-based approaches are not naturally designed for tackling data in non-euclidean space. Hence, more recent models aim to incorporate graph-based methods. GRGNN~\cite{wang2020inductive} was the first to introduce Graph Neural Networks (GNNs)~\cite{liu2023graph,liu2022mlp, yin2023coco, wang2024dusego} into GRN research. It transforms GRN inference into a graph classification task. However, the GNN used by GRGNN does not consider the directionality of gene regulation and stipulates that a gene can only be a transcription factor (TF) or a target gene. In reality, many gene nodes in actual GRNs act as both TFs and target genes. Thus, this method can only infer the existence of regulatory relationships between genes, not the precise regulatory direction. DGCGRN~\cite{wei2024inference} is a GRN inference method based on Directed Graph Convolutional Networks~\cite{tong2020directed}, specifically designed to handle directed graph-structured data. Compared to GRGNN, DGCGRN can better capture the directionality of gene regulation. DGCGRN employs a local enhancement strategy and dynamic updating strategy, generating enhanced features through Conditional Variational Autoencoders~\cite{pol2019anomaly} to address the issue of low-degree nodes, and updates edge weights in each iteration to improve predictive performance. Additionally, DGCGRN incorporates sequence features, providing a more comprehensive inference of gene regulatory relationships for real biological data. However, neither GRGNN nor DGCGRN incorporates prior known gene regulatory networks as prior knowledge. DeepFGRN~\cite{gao2024deepfgrn} effectively reconstructs large-scale and sparse gene regulatory networks by combining correlation analysis with directed graph embedding techniques. This work incorporates known gene regulatory networks to assist in network construction. It is capable of identifying gene regulatory relationships and discerning the types and directions of gene regulation.


When inferring GRNs using directed graph neural networks, genes are treated as nodes and regulatory relationships between TFs and target genes are treated as edges. Although previous methods such as DeepFGRN have incorporated the directed GNN to represent the directionality of GRNs, they neglect the fact that numbers of in/out edges can have a significant gap for each node. In particular,
some genes may regulate the expression of multiple other genes, thus having a higher out-degree. Conversely, if a gene is regulated by multiple factors, it will have a higher in-degree.Such a phenomenon is known as graph with skewed degree distribution~\cite{ke2024duplexdualgatcomplex,wang2024degree}. This challenge is more common in directed graphs, where the separation of in and out neighbours often results in a higher proportion of nodes with a skewed degree distribution compared to undirected graphs. However, existing graph-based bulk GRN inference efforts have not considered this issue, which will affect its prediction performance. In addition, existing approaches use shallow embedding methods such as CNN to capture the correlation between regulator genes and target genes from the bulk gene expression profiles. However, we argue that a more advanced attention mechanism can better capture the complex relations between two genes compared with shallow embedding methods.

To overcome the aforementioned limitations, we introduce a novel approach namedCross-Attention Complex Dual Graph Attention Network Embedding Model (XATGRN). Our XATGRN model is designed to provide a comprehensive understanding of GRNs by predicting the existence of regulatory relationships and determining their directionality and types. In particular, \OURS{} utilizes a cross-attention mechanism to capture the complex interactions reflected in the bulk gene expression profiles of regulator and target genes, thereby enhancing the model's ability to represent these interactions accurately. 

Furthermore, our model employs a sophisticated directed graph representation learning method i.e., DUPLEX~\cite{ke2024duplexdualgatcomplex} to encode the gene-gene relations. Such a DUPLEX method consists of a dual Graph Attention encoder for directional neighbour modelling using the generated amplitude and phase embeddings. 
By leveraging the cross-attention module and the DUPLEX method, our \OURS{} can effectively capture the connectivity and directionality of regulatory interactions within the network and alleviate the issue due to skewed degree distribution in GRNs.

The main contributions of this paper are as follows:
\begin{itemize}
    \item We introduce the XATGRN model, which is capable of predicting the existence, directionality, and type of regulatory relationships in Gene Regulatory Networks (GRNs). This model offers a comprehensive understanding of the intricate mechanisms of gene regulation.
    \item The model utilizes a cross-attention mechanism to focus on the most informative features within bulk gene expression profiles of regulator and target genes, enhancing the model's representational power.
    \item By employing the dual complex graph embedding method, our model generates amplitude and phase embeddings that capture both the connectivity and directionality of regulatory interactions, effectively alleviating the issue due to skewed degree distribution in GRNs.
    \item We conduct extensive experiments on multiple benchmark datasets, demonstrating \OURS{}'s proficiency in uncovering unseen regulatory mechanisms and potential therapeutic targets for complex diseases.
\end{itemize}

\section{Methods}
\subsection{Problem Definition}
In Gene Regulatory Networks (GRNs), regulator genes \( R \) interact with target genes \( T \) to control cellular processes. These interactions can be activating, when the regulator enhances the target's expression, or repressing, when it decreases the expression. The regulatory relationships are directional, flowing from the regulator gene \( R \) to the target gene \( T \). We represent these relationships as directed edges \( \mathbf{e}_{RT} \). The goal is to predict the type of regulation \( \mathbf{r}_{RT} \) between regulator gene \( R \) and target gene \( T \), which can be activation, repression, or non-regulated.

\subsection{Overview of the XATGRN}
\begin{figure*}[htbp]
    \centering
    \includegraphics[width=0.8\textwidth]{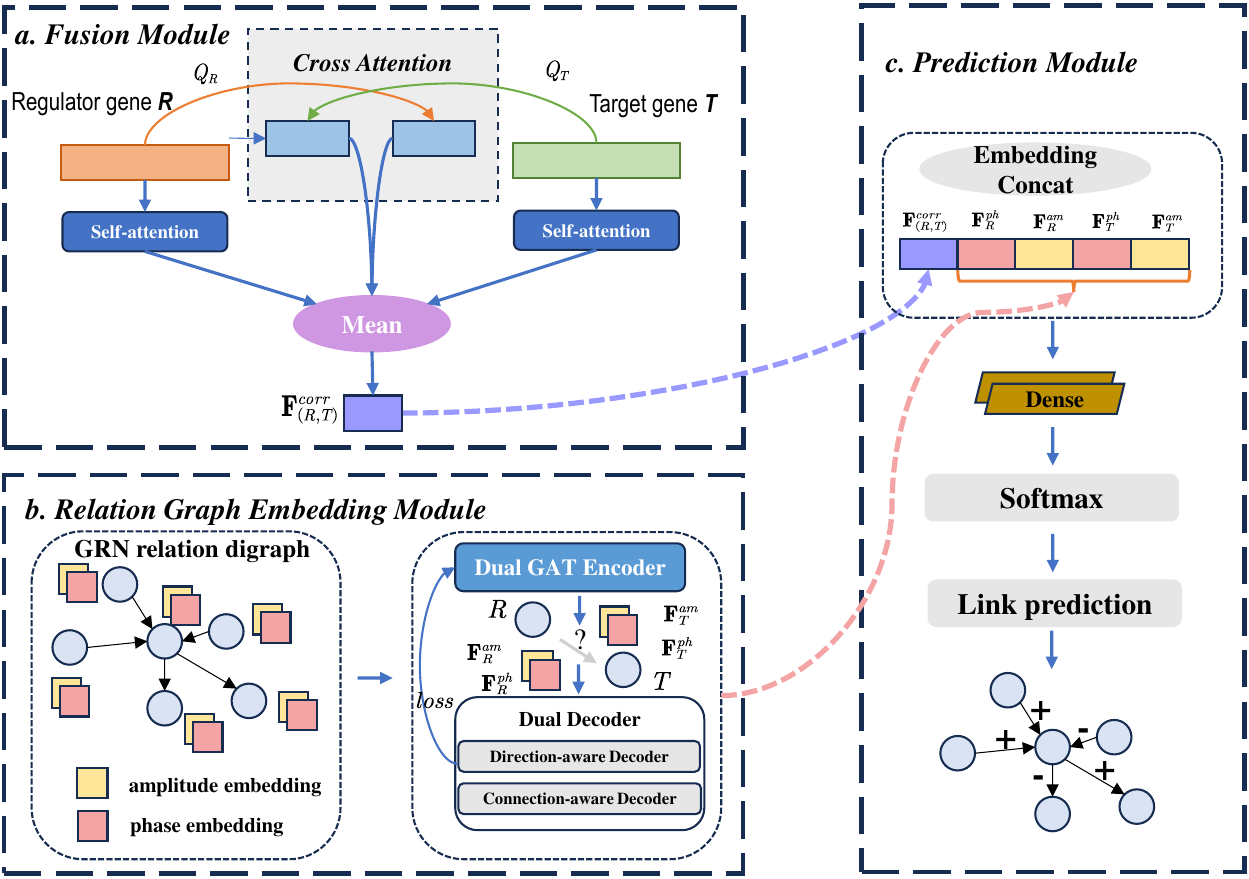}
    \caption{Overview of the XATGRN method.}
    \label{fig:overview}
\end{figure*}

Our Cross-Attention Complex Dual Graph Embedding Model (XATGRN) is designed to infer the regulation types for Gene Regulatory Networks (GRNs). In particular, our \OURS{} can distinguish between the activation type and repression type. Our model operates by treating the GRN inference problem as the link prediction task between regulator genes \( R \) and target genes \( T \). As shown in Figure~\ref{fig:overview} , our model extracts key features from both bulk gene expression data and existing databases that detail prior regulatory associations with regulation types, refining this features through a softmax classifier to predict the regulatory relationships as either activation, repression, or non-regulated interactions.

Initially, the gene expression profiles of regulator-target gene pairs \( (R, T) \) are used by the fusion module(Figure~\ref{fig:overview}a), yielding the fusion embedding vector. This vector encapsulates the gene expression features and the correlation information between the regulator and target genes.
Subsequently, our Relation Graph Embedding Module
The complex embeddings capture  both the connectivity and directionality within the network.


Ultimately, the fusion embedding , along with the complex embeddings of regulator gene \( R \) and target gene \( T \), are concatenated to form a comprehensive feature set in the prediction module(Figure~\ref{fig:overview}c). This aggregated information is then fed into a softmax classifier for predicting the GRN relation.


\subsection{Fusion Module}
\label{sec:correlation_analysis}
Our Fusion Module extracts gene expression features from both the regulator gene \( R \) and the target gene \( T \). This module captures the interactions between the gene pair \( (R, T) \), which are essential for predicting regulatory mechanisms within gene regulatory networks (GRNs), as shown in Figure~\ref{fig:overview}a.

To address the shortcomings of conventional one-dimensional CNNs used by DeepFGRN~\cite{gao2024deepfgrn}, we introduce the Fusion Module.  This module is based on the Cross-Attention Network (CAN), inspired by FusionDTI~\cite{meng2024fusiondtifinegrainedbindingdiscovery}. In particular, CAN enables our model to focus on the most relevant aspects of the gene expressions, which significantly improves its capacity to extract meaningful representations.  

The gene expression data for the regulator gene \( R \) and the target gene \( T \) are processed to generate queries, keys, and values for the cross-attention mechanism, denoted as \( \mathbf{Y}_R \) and \( \mathbf{Y}_T \), respectively. 
The queries, keys, and values for gene \( R \) are represented as \( Q_R, K_R, V_R \) and for gene \( T \) as \( Q_T, K_T, V_T \). The projection matrices \( W_q^R, W_k^R, W_v^R \) and \( W_q^T, W_k^T, W_v^T \) map the gene data into the corresponding representations. Specifically, those matrices are defined as follows:
\begin{align}
Q_R &= Y_R W_q^R, & K_R &= Y_R W_k^R, & V_R &= Y_R W_v^R, \\
Q_T &= Y_T W_q^T, & K_T &= Y_T W_k^T, & V_T &= Y_T W_v^T.
\end{align}


Multi-head self-attention and cross-attention mechanisms are subsequently applied. Notably, each gene retains half of its original self-attention embedding and half of its cross-attention embedding, which allows the model to better handle the intrinsic features of each gene while capturing the complex interactions between them. These embeddings encapsulate the intricate regulatory interactions, thereby enhancing our model's capacity to discern the relationship between the genes.

The embeddings from both the regulator and target genes are then concatenated to form a combined embedding. Such a combined embedding is processed to produce the correlation embedding, which represents the regulatory relationship between the gene pair \( (R, T) \).

The fusion module and the subsequent steps are defined by the following equations:

\begin{align}
R^* = \frac{1}{2}\left[MHA\left(Q_R, K_R, V_R\right) + MHA\left(Q_T, K_R, V_R\right)\right], \\
T^* = \frac{1}{2}\left[MHA\left(Q_T, K_T, V_T\right) + MHA\left(Q_R, K_T, V_T\right)\right], \\
\mathbf{F}_{(R,T)}^{fusion} = \text{Concat}\left(\left(\text{MeanPool}(R^*), \text{MeanPool}(T^*)\right), 1\right),
\end{align}

where the embeddings \( R^* \) and \( T^* \) represent the enhanced representations of the regulator and target genes, respectively. Specifically, $ R^* $ integrates information from both the regulator's self-attention and the cross-attention with the target, while $ T^* $ integrates information from both the target's self-attention and the cross-attention with the regulator. The mean pooling operation is denoted by \( \text{MeanPool} \), and the concatenation operation is denoted by \( \text{Concat} \). The final fusion embedding \( \mathbf{F}_{(R,T)}^{fusion} \) represents the regulatory relationship between the regulator and target genes.

\subsection{Relation Graph Embedding Module}
\label{sec:node_representation}

The Relation Graph Embedding Module addresses the complexity of representing nodes within Gene Regulatory Networks (GRNs). It handles the challenges posed by high-dimensional, sparse, and directed interactions. Specifically, it leverages the skewed degree of genes, which is crucial for differentiating between regulator and target nodes in GRNs.


To effectively embed the nodes in a GRN, we adopt the Complex Dual Graph Embedding approach from the DUPLEX framework~\cite{ke2024duplexdualgatcomplex}. As shown in Figure~\ref{fig:overview}b, this approach generates amplitude and phase embeddings for both regulator and target genes, which encode both the connectivity and directionality of the regulatory interactions.

We model a directed graph (digraph) \( G = (V, E) \), where \( V \) represents the nodes and \( E \) represents the directed edges. Each edge \( (R, T) \in E \) symbolizes a regulatory link from the regulator gene \( R \) to the target gene \( T \). Here, $ R $ and $ T $ are specific instances of genes $ u $ in the graph, where $ R $ acts as a regulator and $ T $ as a target in the regulatory relationship.Our objective is to map each gene \( u \) to a \( d \)-dimensional vector \( \mathbf{x}_u \in \mathbb{C}^{d \times 1} \).

To represent the directionality and connectivity of edges in GRN, our \OURS{} leverage the 
the Hermitian Adjacency Matrix (HAM).
This approach is particularly effective to address the challenge of asymmetric digraphs in GRNs. We use $H$ to denote the HAM, which is defined in polar form as:
\begin{equation}
    H = A_s \odot \exp\left( i \frac{\pi}{2} \Theta \right),
\end{equation}

where \( i \) is the imaginary unit and \( \odot \) represents the Hadamard product. The symmetric binary matrix \( A_s \) is defined as:

\begin{equation}
    A_s(u, v) = \begin{cases}
        1 & \text{if } (u, v) \in E \text{ or } (v, u) \in E, \\
        0 & \text{otherwise}.
    \end{cases}
\end{equation}

The antisymmetric matrix \( \Theta \) contains elements from the set \( \{-1, 0, 1\} \) defined as:

\begin{equation}
    \Theta(u, v) = \begin{cases}
        1 & \text{if } (u, v) \in E, \\
        -1 & \text{if } (v, u) \in E, \\
        0 & \text{otherwise}.
    \end{cases}
\end{equation}

Hence, the HAM's entries \( H(u, v) \), taking values in the set \(\{i, -i, 1, 0\}\), effectively capture the relationships in GRNs, representing four different types of status between $u$ and $v$ including forward, reverse, bidirectional interactions and non-existing.
This representation is particularly suited to GRNs, where regulatory interactions can be both directional and varied in nature. In contrast, the traditional asymmetric adjacency matrix \( A \) necessitates separate entries for \( A(u, v) \) and \( A(v, u) \), each restricted to \(\{0, 1\}\), to encode the same diversity of relationships. The HAM's ability to integrate directionality and connectivity in a single, symmetric matrix offers a more comprehensive and efficient representation, aligning perfectly with the intricate patterns observed in GRNs.


Furthermore, the matrix decomposition \( H = X^{\top} \overline{X} \), where \( X \) is the node embedding matrix, represents the inner product between \( X \) and its complex conjugate \( \overline{X} \). This decomposition facilitates the expression of the node embedding \( \mathbf{x}_u \) in polar form. Specifically, the embedding is written as follows:
\begin{align}
    \mathbf{x}_u &= a_u \odot \exp\left(i \frac{\pi}{2} \theta_u\right), \\
    \overline{\mathbf{x}}_u &= a_u \odot \exp\left(-i \frac{\pi}{2} \theta_u\right).
\end{align}
where \( a_u \) represents the amplitude and \( \theta_u \) the phase of the embedding \( \mathbf{x}_u \). These complex conjugate embeddings \( \mathbf{x}_u \) and \( \overline{\mathbf{x}}_u \) are interpreted as the representations of the regulator and target roles of gene node \( u \), respectively. This joint embedding strategy, in contrast to using separate embeddings for the regulator and target, enables co-optimized learning from both the incoming and outgoing edges of the node \( u \). Such an approach effectively addresses the challenge of the imbalance between in-degrees and out-degrees, which can significantly affect the embedding quality of nodes within Gene Regulatory Networks (GRNs).


\paragraph{Dual GAT Encoder}

Based on HAM, we will introduce a dual encoder architecture that comprises an amplitude encoder, a phase encoder, and a fusion layer.
The encoders refine node embeddings by aggregating information from both incoming and outgoing edges of node\( u \), effectively addressing the issue of skewed degree distribution.

The amplitude encoder employs GAT to aggregate information from both incoming and outgoing edges for each node. This process captures the node's overall connectivity, ensuring that even nodes with varying in-degrees and out-degrees are embedded with high quality in the context of the network's topology:


\begin{equation}
    \mathbf{a}_{u}^{\prime} = \text{ReLU}\left( \sum_{v \in \mathcal{N}(u)} \alpha_{uv}^{(am)} \cdot \mathbf{a}_{v} \right),
\end{equation}

where $\mathcal{N}(u)$ denotes the set of neighboring nodes connected to node $ u $ via either incoming or outgoing edges, \( \alpha_{uv}^{(am)} \) is the attention coefficient for amplitude embedding, and \( \mathbf{a}_{u}^{\prime} \) represents the updated amplitude embedding for node \( u \).

The phase encoder captures the directionality of regulatory relationships by distinguishing between nodes acting as regulators  and targets. The phase embedding is updated similarly using a direction-sensitive attention mechanism:

\begin{equation}
\theta_u' = \text{ReLU} \left( \sum_{v \in \mathcal{N}_{\text{in}}(u)} \alpha_{uv}^{(ph)} \cdot \theta_v - \sum_{v \in \mathcal{N}_{\text{out}}(u)} \alpha_{uv}^{(ph)} \cdot \theta_v \right),
\end{equation}

where $\mathcal{N}_{\text{in}}(u)$ and $\mathcal{N}_{\text{out}}(u)$ denote the sets of in-neighbors and out-neighbors of node $ u $, respectively, and $\alpha_{uv}^{(ph)}$ is the attention coefficient for phase embeddings.

The fusion layer is a critical component that combines the amplitude and phase embeddings, which carry distinct yet complementary information. This layer ensures that the embeddings from both encoders are effectively integrated to capture the comprehensive regulatory interactions within the GRN. The fusion process is designed to balance the contributions from both the amplitude and phase embeddings, thereby enhancing the model's ability to represent complex gene interactions accurately.

The fusion layer operates by combining the information from the amplitude and phase embeddings at each layer of the encoder. This is achieved through a weighted aggregation mechanism, where the attention coefficients dictate the strength of the information exchange between the two types of embeddings. Mathematically, the fusion process for the amplitude embeddings can be formulated as:

\begin{equation}
    \mathbf{F}_u^{am} =  \text{ReLU}\left( \sum_{v \in \mathcal{N}(u)} \alpha_{uv}^{(am)} \cdot \mathbf{a}_{v} + \sum_{v \in \mathcal{N}(u)} \alpha_{uv}^{(ph)} \cdot \mathbf{\theta}_{v} \right),
\end{equation}

where \(\mathbf{F}_u^{am}\) represents the updated amplitude embedding for node \(u\), \(\alpha_{uv}^{(am)}\) and \(\alpha_{uv}^{(ph)}\) are the attention coefficients for amplitude and phase embeddings, respectively.

Similarly, for the phase embeddings, the fusion process is:

\begin{equation}
    \mathbf{F}_u^{ph} =  \text{ReLU}\left( \sum_{v \in \mathcal{N}(u)} \alpha_{uv}^{(ph)} \cdot \mathbf{\theta}_{v} + \sum_{v \in \mathcal{N}(u)} \alpha_{uv}^{(am)} \cdot \mathbf{a}_{v} \right),
\end{equation}

where \(\mathbf{F}_u^{ph}\) represents the updated phase embedding for node \(u\).


\paragraph{Dual Decoders and Loss Functions}

After obtaining the embedding features for each gene node, we introduce  2 parameter-free decoders to reconstruct the Hermitian Adjacency Matrix (HAM) . These decoders are designed to ensure that the embeddings capture both the connectivity and directionality of the regulatory interactions within the GRN.


The direction-aware decoder aims to reconstruct the directionality of regulatory interactions. This task is formulated as a classification problem in GRN , where each edge \( (u, v) \) is assigned probabilities that correspond to its edge type (forward, reverse, bidirectional, or no edge). The predicted edge type is determined by the minimum distance between the estimated matrix element $ \hat{H}(u, v) $ and the possible edge types $ r $:

\begin{equation}
\text{pred. type} = \underset{r}{\text{argmin}} \left( \text{Dist}(\hat{H}(u, v), r) \right),\quad \forall r \in \mathcal{R},
\end{equation}

where $ \text{Dist}(\hat{H}(u, v), r) $ is the distance between $ \hat{H}(u, v) $ and $ r $. and \( \mathcal{R} = \{i, -i, 1, 0\} \) represents the four possible status of edge types.

The probabilities $ P(\hat{H}(u, v) = r) $ are calculated as follows:


\begin{equation}
    P(\hat{H}(u, v) = r) = \frac{\exp(-|\mathbf{x}_u^{\top} \overline{\mathbf{x}}_v - r|)}{\sum_{r' \in \mathcal{R}} \exp(-|\mathbf{x}_u^{\top} \overline{\mathbf{x}}_v - r'|)}, \quad \forall r \in \mathcal{R}.
\end{equation}


The self-supervised direction-aware loss is then defined as:

\begin{equation}
    \mathcal{L}_d = -\sum_{r \in \mathcal{R}} \sum_{H(u, v) = r} \log P(\hat{H}(u, v) = r).
\end{equation}


The connection-aware decoder is designed to reconstruct the binary presence of connections between genes, which are encoded in the amplitude embeddings. It models the connection probability for an edge \( (u, v) \) as:

\begin{equation}
    P(\hat{A}_s(u, v) = 1) = \sigma(a_u^{\top} a_v),
\end{equation}

where \( \sigma \) is the sigmoid function and \( \hat{A}_s \) is the estimated connection matrix, and $\hat{A}_s$ represent the probability that a connection exists between genes $u$ and $v$, allowing the model to capture uncertainty in the network structure. The connection-aware loss function \( \mathcal{L}_c \) adheres to the same negative log-likelihood minimization principle as the direction-aware loss:

\begin{equation}
    \mathcal{L}_c = -\sum_{u, v} \log P(\hat{A}_s(u, v) = A_s(u, v)).
\end{equation}


The total loss function combines the direction-aware and connection-aware losses to ensure that the model learns both the connectivity and directionality of the regulatory interactions. The total loss is given by:

\begin{equation}
    \mathcal{L}_{total} = \mathcal{L}_d + \lambda \mathcal{L}_c,
\end{equation}

where \(\lambda\) is a hyperparameter that controls the weight of the connection-aware loss. 


By leveraging the fusion module and the Relation Graph Embedding Module, our XATGRN can effectively capture the connectivity and directionality of regulatory interactions within the network and alleviate the issue due to skewed degree distribution in GRNs.

\subsection{Prediction Module}

The prediction module in our model is designed to leverage the embeddings generated by the Fusion Module and the Relation Graph Embedding Module to make accurate predictions about the regulatory relationships within the Gene Regulatory Network (GRN). As shown in Figure~\ref{fig:overview}c, This module integrates the complex embeddings of genes and employs a series of neural network layers to classify the interactions between gene pairs $(R, T)$.

For each sample, the module processes the feature and label data for a gene pair $(R, T)$, where $ R $ is the regulator gene and $ T $ is the target gene. The features from the Fusion Module are denoted as $ \mathbf{F}_{(R,T)}^{fusion} $, while the features from the Relation Graph Embedding Module include the amplitude and phase embeddings for both genes: $ \mathbf{F}_R^{am} $, $ \mathbf{F}_T^{am} $, $ \mathbf{F}_R^{ph} $, and $ \mathbf{F}_T^{ph} $.


These embeddings are concatenated to form a comprehensive feature vector:

\begin{align}
    \mathbf{F} &= \text{concat}(\mathbf{F}_{(R,T)}^{fusion}, \mathbf{F}_R^{am}, \mathbf{F}_T^{am}, \mathbf{F}_R^{ph}, \mathbf{F}_T^{ph}).
\end{align}

The concatenated feature vector $\mathbf{F}$ is first processed through a 1-dimensional convolutional layer with batch normalization (BN) and max pooling, followed by ReLU activation to extract and refine spatial features:

\begin{align}
    x_1 &= \text{ReLU}(\text{MaxPool}(\text{BN}(\text{Conv1D}(\mathbf{F})))),
\end{align}

where the output is $ x_1 $, and the vector $ x_1 $ is pass to the global average pooling layer:

\begin{align}
    x_2 &= \text{GlobalAvgPool}(x_1),
\end{align}

where the resulting vector is denoted as $ x_2 $.This step condenses the feature map into a fixed-size vector that captures the essential information for classification. 

Subsequently, the flattened vector $ x_2 $ is passed through two fully connected layers(\(\text{FC}_1\) and \(\text{FC}_2\)) with dropout for regularization, where a dropout rate of \(p=0.3\) is applied:

\begin{align}
    x_3 &= \text{FC}_2(\text{Dropout}(\text{ReLU}(\text{FC}_1(x_2)), p=0.3)),
\end{align}

where $x_3$ represents the output of the second fully connected layer, and $x_3$ is then passed through a softmax function to produce the final classification probabilities over $C$ classes:

\begin{align}
    \text{output} &= \text{softmax}(x_3)
\end{align}

To address class imbalance, we employ a weighted cross-entropy loss function \(\mathcal{L}\). The weights \(w_c\) for each class \(c\) are inversely proportional to their frequency in the dataset:

\begin{align}
    w_c &= \frac{N_{\text{total}}}{N_c},
\end{align}
where \(N_{\text{total}}\) is the total number of samples in the dataset, and \(N_c\) is the number of samples in class \(c\).

The loss function \(\mathcal{L}\) is then defined as:

\begin{align}
    \mathcal{L} &= -\sum_{c=1}^{C} w_c \cdot y_c \cdot \log(\hat{y}_c),
\end{align}
where \(y_c\) is the true label for class \(c\), and \(\hat{y}_c\) is the predicted probability for class \(c\).

Our  model optimizes this loss function using the Adam optimizer, with an exponential learning rate scheduler to ensure stable and efficient convergence.

\section{Dataset and Experiment}

\subsection{Datasets}



To examine the performance of our XATGRN model, we use the FGRN benchmark, which is introduced in the DeepFGRN~\cite{gao2024deepfgrn}. The benchmark collects bulk gene expression data and prior regulatory gene pairs with regulation types across 9 distinct datasets. These datasets include the DREAM5 challenge network1, E.coli under 4 different stress conditions (cold, heat, lactose, and oxidative stress), and 4 types of human diseases (Covid-19, breast cancer, lung cancer, and liver cancer). These datasets are particularly relevant for studying disease mechanisms from a gene regulatory perspective due to their significant implications in understanding disease pathology.

\begin{table}[h!]
\centering
\begin{tabular}{cccccc}
    \toprule
    Species & Datasets & numG & dimG & numA & numR \\
    \midrule
    DREAM5 & network1 & 1643 & 805 & 2236 & 1776 \\
    E.coli & cold stress & 2205 & 24 & 2070 & 2034 \\
    E.coli & heat stress & 2205 & 24 & 2070 & 2034 \\
    E.coli & lactose stress & 2205 & 12 & 2070 & 2034 \\
    E.coli & oxidative stress & 2205 & 33 & 2070 & 2034 \\
    Human & COVID-19 & 2478 & 42 & 6452 & 1888 \\
    Human & Breast cancer & 2478 & 24 & 6452 & 1888 \\
    Human & Liver cancer & 2478 & 10 & 6452 & 1888 \\
    Human & Lung cancer & 2478 & 130 & 6452 & 1888 \\
    \bottomrule
    \end{tabular}
    \caption{Summary of benchmark datasets statics.}
    \label{dataset static}
\end{table}

The detailed statistics of these datasets are summarized in Table~\ref{dataset static}. The table includes the number of genes (numG), the dimension of gene expression data (dimG), and the number of regulatory associations for known activation types (numA) and repression types (numR).

\begin{figure}[htbp]
    \centering
    \includegraphics[width=\linewidth]{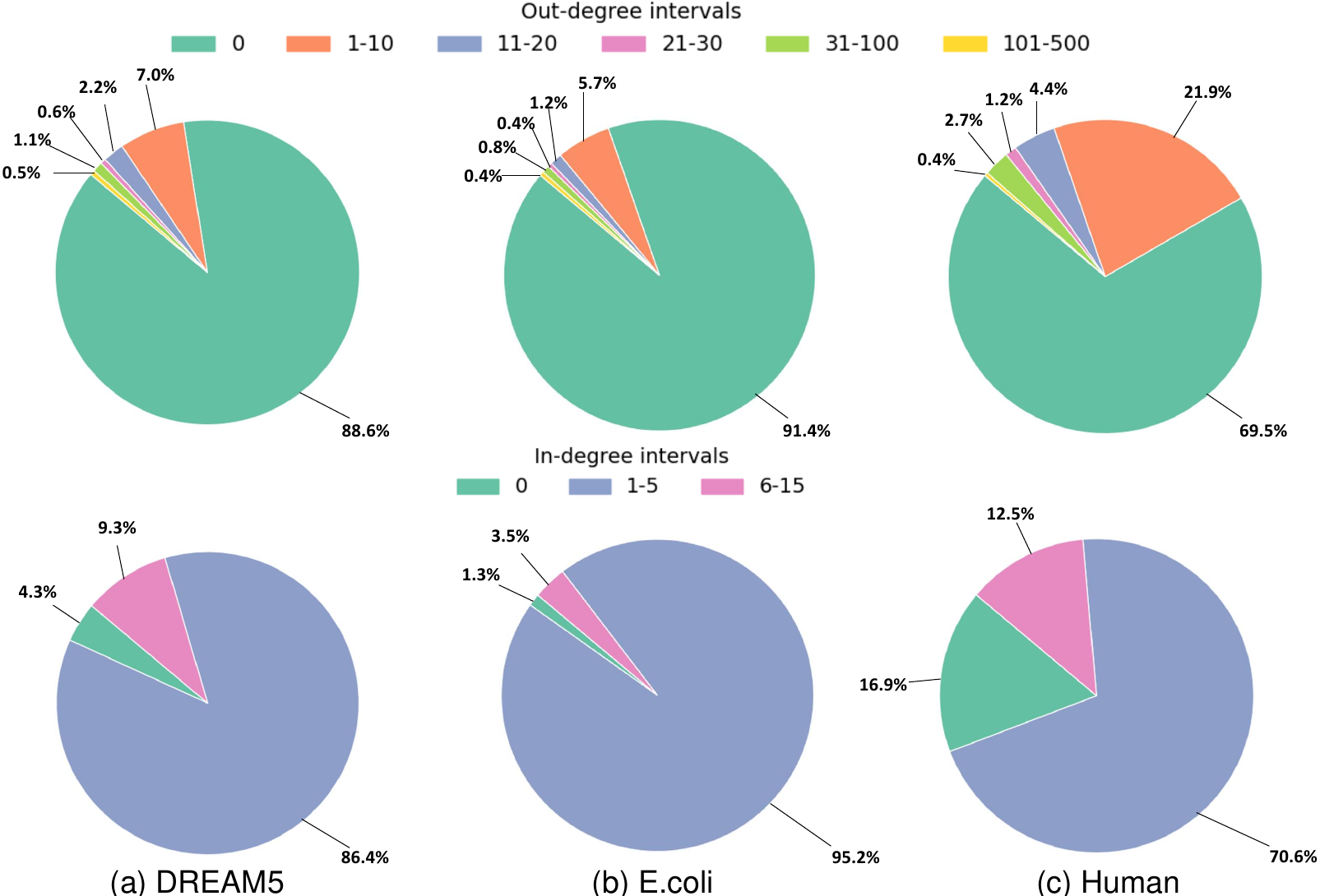}
    \caption{Visualization of the in-degree and out-degree distribution in DREAM5, Human and E.coli datasets}
    \label{fig:skewed}
\end{figure}

To illustrate the skewed degree distribution in these datasets, we provide a visualization of the in-degree and out-degree distributions for the DREAM5, Human, and E.coli datasets (Figure~\ref{fig:skewed}). This visualization highlights the significant variation in the number of incoming and outgoing edges for different genes, a characteristic that poses a challenge for traditional graph embedding methods.

\subsection{Experiment Setting}
To better evaluate the performance of the model, we adapt the 10 times of the 5-fold cross-validation. For each fold, we calculate the mean of Area Under the Receiver Operating Characteristic curve (AUC), precision, recall, and F1-score. These metrics provide a comprehensive assessment of the model's ability to accurately predict both the presence and type of regulatory interactions within gene networks.
For Fusion module,we use pytorch for implement. we employed the Adam optimizer with a learning rate of 0.001 to update the model parameters during training.
For Relation Graph Embedding Module, we use DGL library and Pytorch for implementation. with a learning rate of 1e-3. We set the hidden dim to 128 and the dropout rate to 0.5. We select the optimal hyperparaneter for initial loss weight \(\lambda\)  in {0.1,0.3} and the decay rate $q$ in {0,1e-4,1e-2}. To demonstrate the effectiveness of our model, we compare \OURS{} with state-of-the-art GRN inference models, including CNNGRN~\cite{CNNGRN}, DGCGRN~\cite{DGCGRN}, and DeepFGRN~\cite{dong2024deep}. 

\subsection{Experiment Result}

\begin{table*}[htbp]
\centering
\small
\begin{tabular}{lcccccccccccc}
\toprule
Model & \multicolumn{4}{c}{DREAM5 network1} & \multicolumn{4}{c}{E.coli cold stress} & \multicolumn{4}{c}{E.coli heat stress} \\ 
\cmidrule(lr){2-5} \cmidrule(lr){6-9} \cmidrule(lr){10-13}
 & AUC & Recall & F1 & Precision & AUC & Recall & F1 & Precision & AUC & Recall & F1 & Precision \\
\midrule
CNNGRN & 0.7682& 0.6304& 0.5932& 0.5549& 0.8443& 0.5941& 0.5538& 0.5183& 0.8494& 0.5811& 0.5488& 0.5236\\
DGCGRN & 0.8751& 0.7773& 0.7786& 0.7823& 0.8121& 0.7198& 0.7209& 0.7244& 0.8140& 0.7209& 0.7219& 0.7250\\
DeepFGRN  & 0.9255& 0.7993& 0.8000& 0.8039& 0.9196& 0.7935& 0.7943& 0.7975& 0.9180& 0.7896& 0.7907& 0.7938\\
XATGRN & \textbf{0.944}7& \textbf{0.8392}& \textbf{0.8410}& \textbf{0.8448}& \textbf{0.9217}& \textbf{0.8167}& \textbf{0.8178}& \textbf{0.8219}& \textbf{0.9216}&\textbf{ 0.8163}&\textbf{ 0.8174}& \textbf{0.8208}\\
\midrule
& \multicolumn{4}{c}{E.coli lactose stress} & \multicolumn{4}{c}{E.coli oxidative stress} & \multicolumn{4}{c}{Human COVID-19} \\ 
\cmidrule(lr){2-5} \cmidrule(lr){6-9} \cmidrule(lr){10-13}
 & AUC & Recall & F1 & Precision & AUC & Recall & F1 & Precision & AUC & Recall & F1 & Precision \\
\midrule
CNNGRN & 0.8386& 0.6129& 0.5962& 0.5603& 0.8402& 0.5983& 0.5668& 0.5381& 0.8629& 0.7418& 0.7134& 0.6857\\
DGCGRN & 0.8240& 0.7299& 0.7308& 0.7346& 0.8203& 0.7269& 0.7280& 0.7313& 0.8146& 0.7083& 0.7175& 0.7330\\
DeepFGRN  & 0.9210& 0.7962& 0.7969& 0.7999& 0.9175& 0.7913& 0.7923& 0.7952& 0.9002& \textbf{0.7943}& 0.7849& 0.7833\\
XATGRN & \textbf{0.9208}& \textbf{0.8161}&\textbf{ 0.8171}& \textbf{0.8209}&\textbf{ 0.9228}& \textbf{0.8180}& \textbf{0.8191}&\textbf{ 0.8233}& \textbf{0.9105}& 0.7916&\textbf{ 0.8018}& \textbf{0.8192}\\
\midrule
& \multicolumn{4}{c}{Human Breast cancer} & \multicolumn{4}{c}{Human Liver cancer} & \multicolumn{4}{c}{Human Lung cancer} \\ 
\cmidrule(lr){2-5} \cmidrule(lr){6-9} \cmidrule(lr){10-13}
 & AUC & Recall & F1 & Precision & AUC & Recall & F1 & Precision & AUC & Recall & F1 & Precision \\
\midrule
CNNGRN & 0.8619& 0.7486& 0.6811& 0.6532& 0.8594& 0.7822& 0.7468& 0.7183& 0.8438& 0.7203& 0.6648& 0.6249\\
DGCGRN & 0.8137& 0.7087& 0.7168& 0.7297& 0.8198& 0.7160& 0.7249& 0.7369& 0.8199& 0.7183& 0.7262& 0.7390\\
DeepFGRN & 0.9013& \textbf{0.7969}& 0.7852& 0.7839& 0.8966& 0.7984& 0.7875& 0.7853& 0.9049& \textbf{0.8047}& 0.7959& 0.7938\\
XATGRN & \textbf{0.9086}& 0.7923& \textbf{0.8011}&\textbf{ 0.815}3& \textbf{0.9150}& \textbf{0.8002}& \textbf{0.8090}&\textbf{ 0.8236}& \textbf{0.9159}& 0.8028& \textbf{0.8114}& \textbf{0.8257}\\
\bottomrule
\end{tabular}
\caption{Average results over 10 times of 5-fold cross-validation for model performance comparisons across 9 datasets (DREAM5 network1, E.coli cold stress, E.coli heat stress, E.coli lactose stress, E.coli oxidative stress, Human COVID-19, Human Breast cancer, Human Liver cancer, Human Lung cancer)}
\label{main table}
\end{table*}

\begin{figure*}[htbp]
    \centering
    \includegraphics[width=0.9\linewidth]{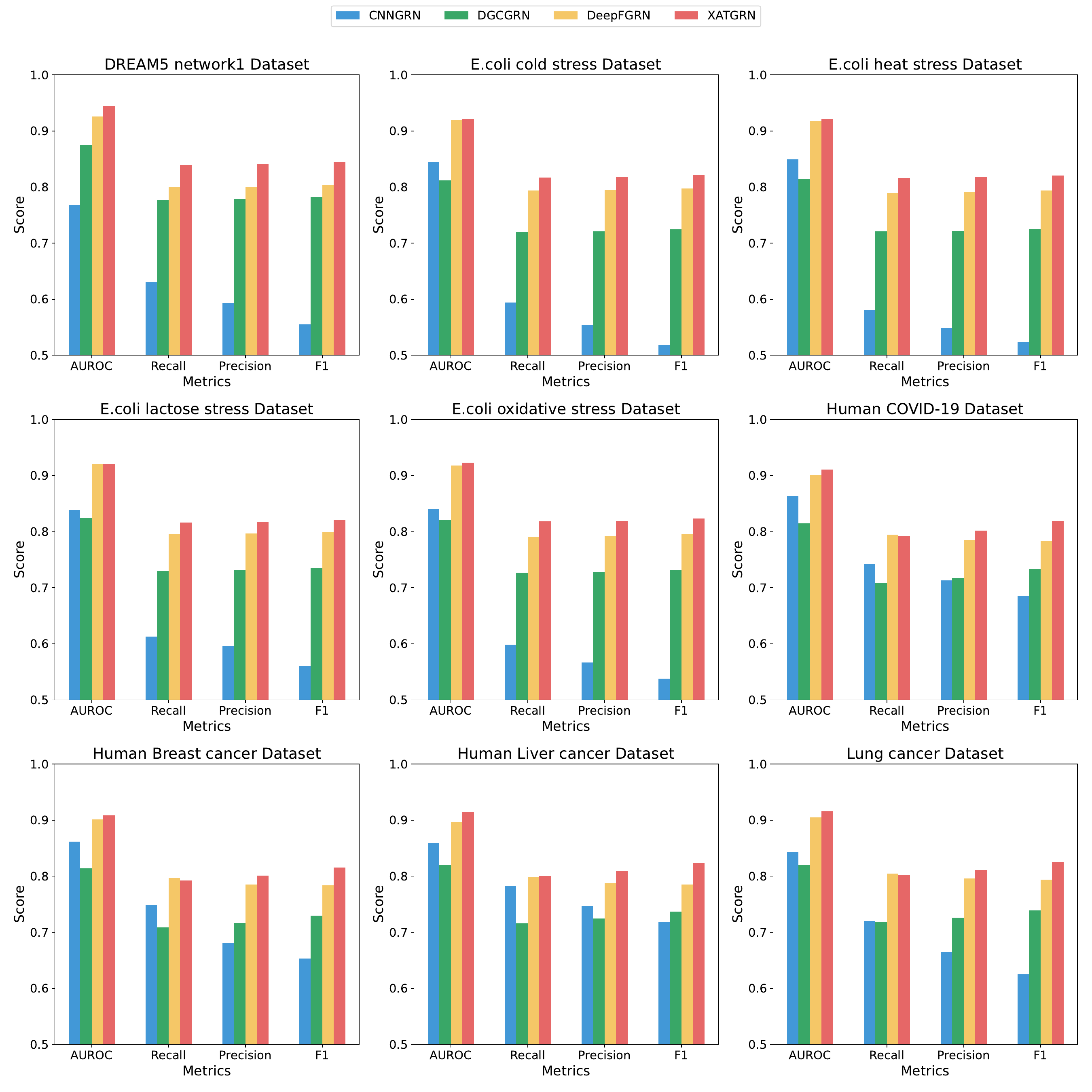}
    \caption{Average results over 10 times of 5-fold cross-validationfor model performance comparisons across 9 datasets }
    \label{fig:main fig}
\end{figure*}


As shown in Table~\ref{main table} and Figure~\ref{fig:main fig}, our XATGRN model consistently outperforms state-of-the-art models across all datasets. In particular, CNNGRN focuses on extracting and reconstructing features from bulk gene expression data but disregards the original structure of the gene regulatory network (GRN). While DGCGRN and DeepFGRN construct directed graph embeddings for GRN inference, they fail to address the challenge posed by skewed degree distributions, which can lead to suboptimal performance, particularly for genes with significant disparities between in-degree and out-degree variances.
Our XATGRN achieves the highest AUC, recall, F1-score, and precision across the DREAM5 network1 and all four E.coli datasets. These results demonstrate that XATGRN effectively captures the complex regulatory interactions within gene networks and accurately predicts both the presence and types of regulatory relationships. Notably, XATGRN's robust performance highlights its ability to handle the skewed degree challenge more effectively compared to DeepFGRN and DGCGRN.
However, it is worth noting that in certain cases, such as the human COVID-19, breast cancer, and lung cancer datasets, the recall of XATGRN is slightly lower than that of DeepFGRN. 
This observation indicates that it may be necessary for such complex datasets to strike a better balance between addressing the skewed degree problem and optimizing source-target embeddings. In conclusion, our \OURS{} model constantly and consistently outperform competitive baselines for inferencing GRNs from different types of gene expression data.

\subsubsection{Ablation Study}
\begin{figure*}[hbp]
    \centering
    \includegraphics[width=0.9\linewidth]{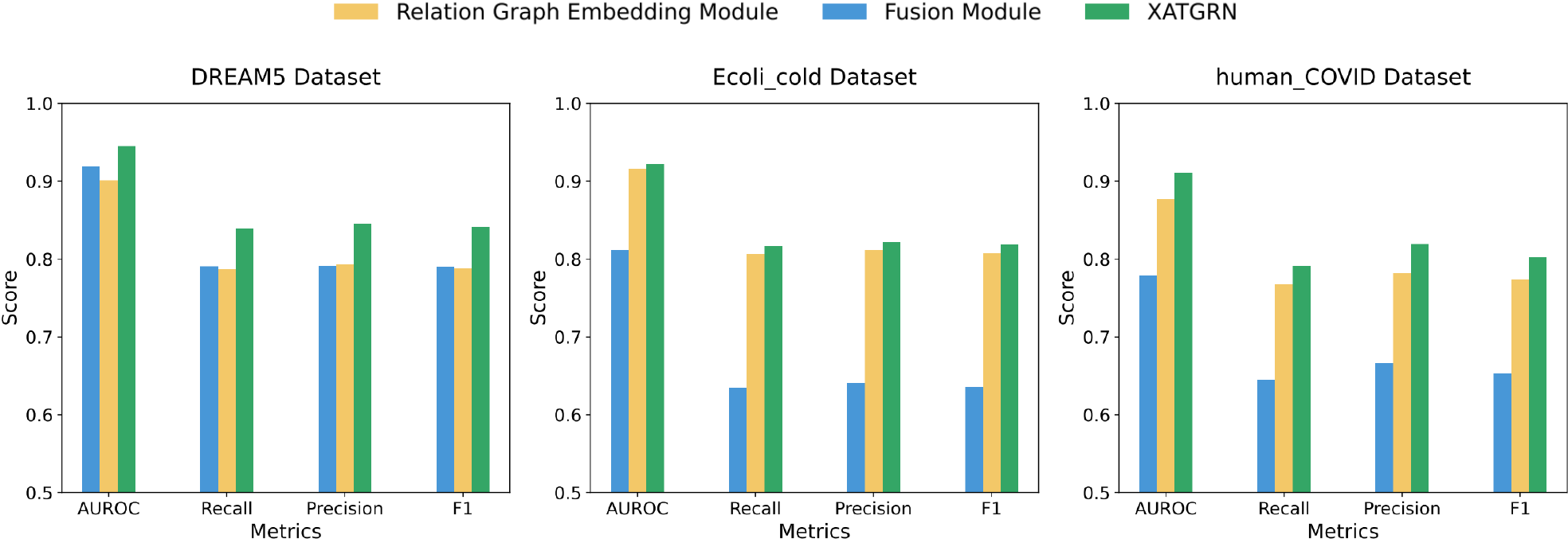}
    \caption{Ablation study in DREAM5 Ecoli cold and human COVID datasets}
    \label{fig:ablation}
\end{figure*}
To evaluate the contribution of each module in the XATGRN model, we conducted an ablation study by systematically varying the inclusion of key components. The study was performed on three datasets: DREAM5 Network1, E.coli cold stress, and the Human COVID-19 dataset. We compared three different setups to assess the impact of the Fusion Module and the Relation Graph Embedding Module on the model's performance. Specifically, these setups included a full \OURS{} model with both modules, a model with only the Relation Graph Embedding Module, and a model with only the Fusion Module. Detailed results are 
visualized in Figure~\ref{fig:ablation}.

The full \OURS{} model, which includes both the Fusion Module and the Relation Graph Embedding Module, achieved the highest performance across all metrics. This configuration effectively captures the complex interactions within the Gene Regulatory Network (GRN), handling the skewed degree distribution of genes while preserving accuracy and robustness.

When only the Relation Graph Embedding Module was included, the model's performance dropped slightly. These reductions highlight the importance of the Fusion Module in enhancing the model's discriminative power. By focusing on the most relevant gene expression features and the correlations between regulator and target genes, the Fusion Module improves the model's ability to accurately distinguish activation, repression, and non-regulated interactions.

Conversely, when only the Fusion Module was included, the model's performance also declined. This underscores the critical role of the Relation Graph Embedding Module in preserving the model's accuracy and robustness. The Relation Graph Embedding Module captures the connectivity and directionality of interactions within the GRN, allowing the model to effectively handle the skewed degree distribution of genes.

These findings demonstrate that the combination of the  Relation Graph Embedding module and the Fusion module greatly enhances the model's ability to accurately predict regulatory interactions. Both components are indispensable for the model's success.

\section{Case study}
\begin{figure}[htbp]
    \centering
    \includegraphics[width=1\linewidth]{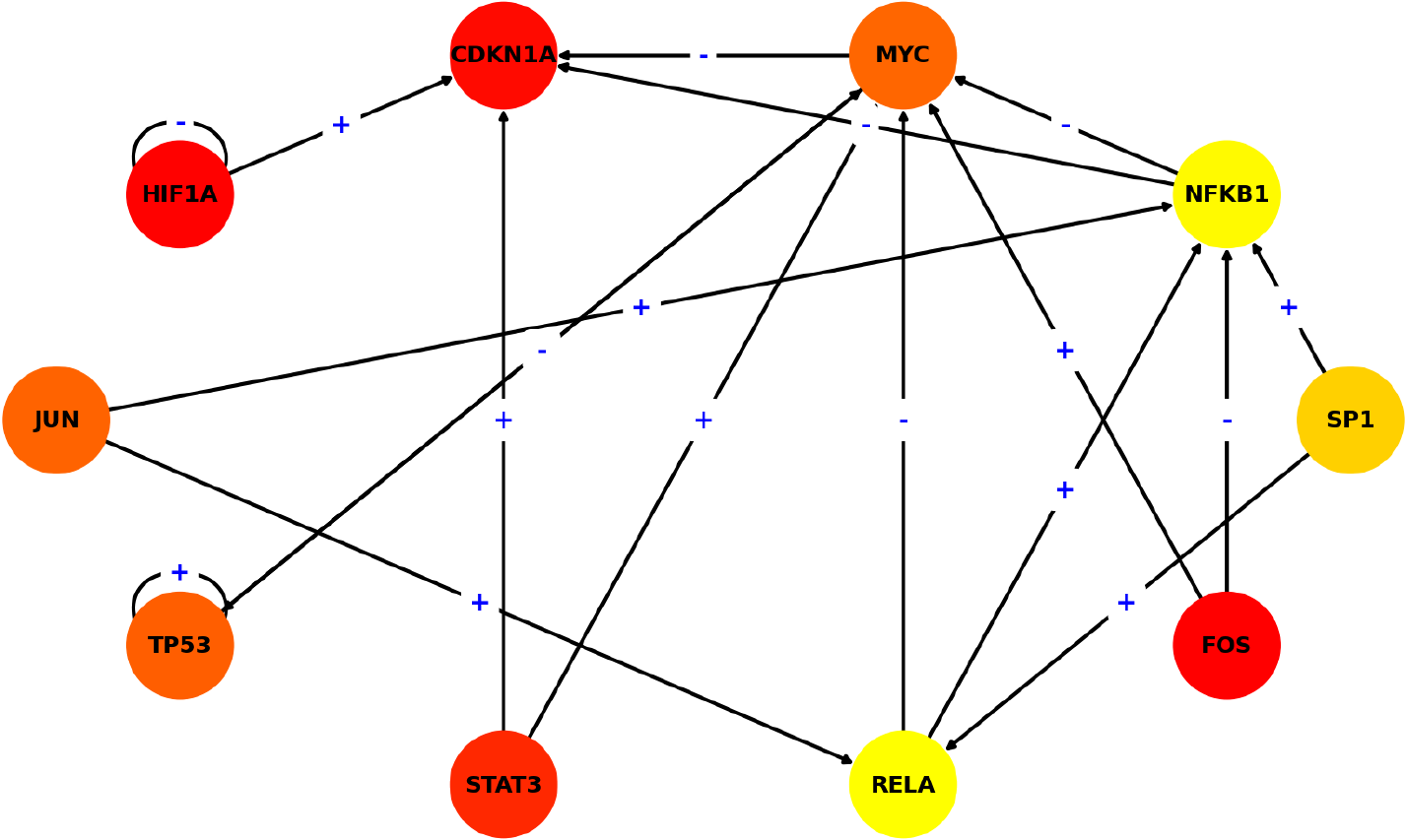}
    \caption{Hub gene of the breast cancer}
    \label{fig:Breast_graph}
\end{figure}

To validate the biological significance  of \OURS{}, we reconstructed GRN using brest cancer data and employed in-depth analysis including prediction of biomarkers and enrichment analysis of potential therapeutic drugs. The constructed breast cancer GRN contains 2,478 genes and 8,772 relationships. After reconstructing GRN, we selected ten hub genes with the highest degree (Figure~\ref{fig:Breast_graph}).

The hub genes in Figure~\ref{fig:Breast_graph} have been fully validated through literature review. Both RELA and NFKB1 are important members of the nuclear factor kappa-B family. In breast cancer, the abnormal activation of the kappa-B signaling pathway is closely associated with the occurrence, development, invasion, and metastasis of tumors~\cite{karin2002nf, biswas2001nuclear}. SP1 is closely related to the staging, invasive potential, and survival rates of breast cancer, and high levels of SP1 often indicate poor prognosis for patients~\cite{zhao2018tgf}.
MYC is a key regulator of cell growth, proliferation, metabolism, differentiation, and apoptosis, and its deregulation contributes to breast cancer development and progression, associated with poor outcomes~\cite{xu2010myc}. The transcription factor Jun is closely associated with metastasis and prognosis in breast cancer, acting as both a suppressor and oncogene\cite{zhu2022transcription}. TP53 mutation, frequently occurring in triple-negative breast cancer (TNBC), enhances the correlation between the high-MYC and low-TXNIP gene signature and death from breast cancer~\cite{borresen2003tp53}. STAT3 plays a crucial role in the regulation of cancer hallmarks in breast cancer, including angiogenesis, metabolism, and invasion, and is involved in tamoxifen resistance~\cite{mirzaei2023hypoxia}. The cytoplasmic localization of CDKN1A/p21 is predominantly associated with cancer, where it serves to promote tumorigenesis and inhibit apoptosis in breast cancer cell lines~\cite{wei2015expression}. Hypoxia inducible factor-1$ \alpha $ (HIF-1$ \alpha $) is crucial in the regulation of cancer hallmarks in breast cancer, including angiogenesis, metabolism, and invasion~\cite{mirzaei2023hypoxia}. Lastly, FOS is downregulated in breast cancer tissues and cells, and its overexpression restrains the malignant phenotypes of breast cancer cells~\cite{chang2023targeting}.

\begin{figure}[htbp]
    \centering
    \includegraphics[width=1\linewidth]{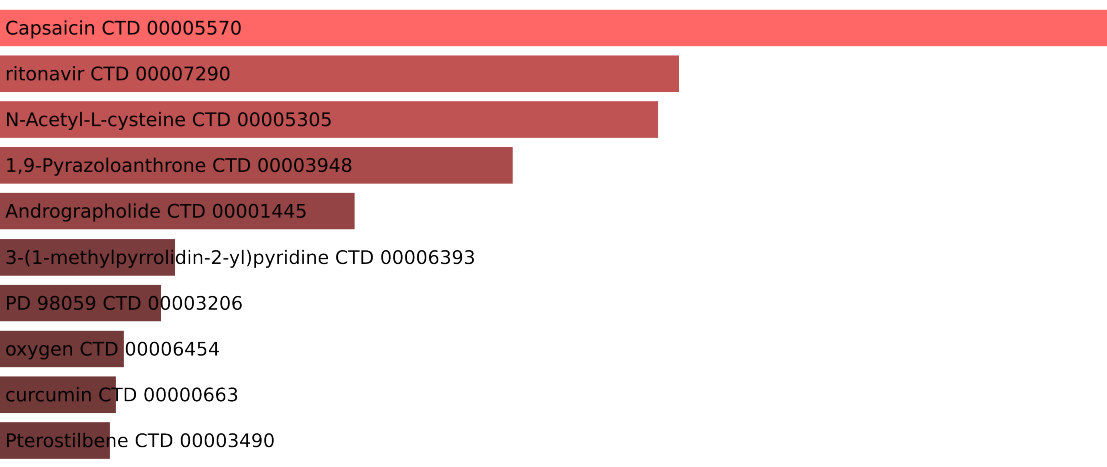}
    \caption{Drug enrichment analysis of breast cancer.}
    \label{fig:drug}
\end{figure}
Furthermore, we performed drug enrichment analysis on top ten 
hub genes of breast cancer predicted by \OURS{}. Figure~\ref{fig:drug} shows the top 10 enriched potential drugs based 
on DsigDB obtained through hub genes. It has been confirmed 
that seven out of the 10 drugs in figure~\ref{fig:drug} may be used for 
the treatment of breast cancer. Capsaicin, known for its activation of the TRPV1 receptor, has shown potential in inhibiting breast cancer cell growth by inducing apoptosis~\cite{chen2021capsaicin}. Ritonavir, typically used as an HIV protease inhibitor, is being studied for its potential to boost the effectiveness of breast cancer chemotherapy~\cite{hendrikx2016ritonavir}. N-Acetyl-L-cysteine (NAC), with its antioxidant properties, shows promise in mitigating chemotherapy side effects and may play a role in breast cancer management~\cite{cheng2013effects}. Andrographolide has been found to suppress breast cancer progression by inhibiting COX-2 expression and angiogenesis, affecting p300 signaling and the VEGF pathway~\cite{peng2018andrographolide}. PD 98059, a MEK inhibitor, reduces the invasive capabilities of breast cancer cells by disrupting the MAPK signaling pathway~\cite{normanno2006mek}. Hyperbaric oxygen therapy is being explored as a supportive measure to improve the outcomes of radiotherapy for breast cancer~\cite{batenburg2021impact}. Curcumin, a component of turmeric, has shown potential in breast cancer treatment due to its anti-tumor, anti-oxidative, and anti-inflammatory properties, coupled with its low toxicity and high safety profile~\cite{hu2018curcumin}.  

\section{Conclusion and Discussion}

In this paper, we have introduced the Cross-Attention Complex Dual Graph Attention Network Embedding Model (XATGRN) for Gene Regulatory Network (GRN) inference. This model addresses several critical challenges in GRN prediction, including the accurate representation of gene regulatory interactions, the handling of skewed degree distributions, and the effective capture of complex gene-gene relationships. By incorporating a cross-attention mechanism, XATGRN enhances the model’s ability to predict not only the presence of regulatory relationships but also their directionality and specific types—such as activation or repression.

Our results show that XATGRN consistently outperforms state-of-the-art models across multiple datasets. The cross-attention mechanism allows XATGRN to focus on the most relevant features from bulk gene expression data, while our relation graph embedding module effectively captures both connectivity and directionality within the GRN, even in the presence of imbalanced node degrees. This combination of strategies enables XATGRN to overcome the limitations of existing models, making it more robust and applicable in real-world biological contexts. The strong performance of XATGRN across diverse datasets emphasizes its robustness and generalizability, positioning it as a promising tool for exploring GRNs in a variety of biological contexts.

Extensive experiments on benchmark datasets underscore the model’s effectiveness in uncovering previously unknown regulatory mechanisms and its potential to identify novel therapeutic targets for complex diseases. Our XATGRN model provides a comprehensive and powerful framework for advancing our understanding of gene regulatory networks, offering a valuable approach for both basic and applied biological research.

In conclusion, our XATGRN represents a significant step forward in GRN inference, providing a robust and accurate framework for studying gene regulatory mechanisms. By effectively managing skewed degree distributions and leveraging advanced attention mechanisms, XATGRN serves as a powerful tool for uncovering regulatory interactions and identifying potential therapeutic targets.


\bibliographystyle{unsrt}
\bibliography{reference}
\end{document}